\documentclass[aps,prd,twocolumn,showpacs,groupedaddress]{revtex4}  
\usepackage{graphicx}  
\usepackage{dcolumn}   
\usepackage{bm}        
\usepackage{amssymb}   
\usepackage{amsmath}
\usepackage{multirow}

\def\met{\mbox{${\hbox{$E$\kern-0.6em\lower-.1ex\hbox{/}}}_T$}}

\begin{document}

\hspace{5.2in} \mbox{FERMILAB-PUB-09-380-E}

\title{Combined measurements of anomalous charged trilinear gauge-boson couplings from diboson production 
in {\boldmath$p\bar{p}$} collisions at {\boldmath$\sqrt{s}=1.96$} TeV}
%
%
\author{V.M.~Abazov$^{37}$}
\author{B.~Abbott$^{75}$}
\author{M.~Abolins$^{65}$}
\author{B.S.~Acharya$^{30}$}
\author{M.~Adams$^{51}$}
\author{T.~Adams$^{49}$}
\author{E.~Aguilo$^{6}$}
\author{M.~Ahsan$^{59}$}
\author{G.D.~Alexeev$^{37}$}
\author{G.~Alkhazov$^{41}$}
% ...

\affiliation{\vspace{0.1 in}(The D\O\ Collaboration)\vspace{0.1 in}}
% ...
\affiliation{$^{6}$University of Alberta, Edmonton, Alberta, Canada;
                Simon Fraser University, Burnaby, British Columbia, Canada;
                York University, Toronto, Ontario, Canada and
                McGill University, Montreal, Quebec, Canada}
% ...
\affiliation{$^{30}$Tata Institute of Fundamental Research, Mumbai, India}
% ...
\affiliation{$^{37}$Joint Institute for Nuclear Research, Dubna, Russia}
% ...
\affiliation{$^{41}$Petersburg Nuclear Physics Institute,
                St. Petersburg, Russia}
% ...
\affiliation{$^{49}$Florida State University, Tallahassee, Florida 32306, USA}
% ...
\affiliation{$^{51}$University of Illinois at Chicago,
                Chicago, Illinois 60607, USA}
% ...
\affiliation{$^{59}$Kansas State University, Manhattan, Kansas 66506, USA}
% ...
\affiliation{$^{65}$Michigan State University,
                East Lansing, Michigan 48824, USA}
% ...
\affiliation{$^{75}$University of Oklahoma, Norman, Oklahoma 73019, USA}
% ...
\date{July 28, 2009}

\begin{abstract} 
  We present measurements of the anomalous $WW\gamma$ and $WWZ$ trilinear gauge couplings from a combination of four diboson production and decay channels using data collected by the D0 detector at the Fermilab Tevatron Collider. These results represent the first high statistics combination of limits across different diboson production processes at the Tevatron and use data corresponding to an integrated luminosity of approximately 1~fb$^{-1}$. When respecting $SU(2)_L\otimes U(1)_Y$ symmetry, we measure central values and 68\% C.L. allowed intervals of $\kappa_{\gamma}=1.07^{+0.16}_{-0.20}$, $\lambda =0.00^{+0.05}_{-0.04}$ and $g_{1}^{Z}=1.05 \pm 0.06$. We present the most stringent measurements to date for the $W$ boson magnetic dipole and electromagnetic quadrupole moments of $\mu_W=2.02^{+0.08}_{-0.09} \, (e/2M_W)$ and $q_W=-1.00\pm0.09 \, (e/M^2_W)$, respectively.
 \end{abstract}

\maketitle

The gauge theory of electroweak interactions contains a striking feature. In quantum electrodynamics, the photons carry no electric charge and thus lack photon-to-photon couplings and do not self-interact. In contrast, the weak vector bosons carry weak charge and do interact amongst themselves through trilinear and quartic gauge boson vertices.

The most general $WW\gamma$ and $WWZ$ interactions can be described~\cite{lagrangian,HWZ} using a Lorentz invariant effective Lagrangian that contains fourteen dimensionless couplings, seven each for $WW\gamma$ and $WWZ$.  Assuming electromagnetic gauge invariance and $CP$ conservation reduces the number of independent couplings to five (electromagnetic gauge invariance requires $g^\gamma_1=1$), and the Lagrangian takes the form:

\begin{equation}
 \label{eq-lag}
 \begin{split}
 \frac{\mathcal{L}_{WWV}}{g_{WWV}} & = ig^V_1(W^\dagger_{\mu\nu}W^{\mu}V^{\nu} -
 W^\dagger_{\mu}V_{\nu}W^{\mu\nu})  \nonumber \\  
 & {} + i \kappa_V W^{\dagger}_\mu W_\nu
 V^{\mu\nu} + \frac{i\lambda_V}{M^2_W} W^\dagger_{\lambda\mu}W^\mu{}_\nu
 V^{\nu\lambda} \\
 \end{split}
 \end{equation}

\noindent where $W^{\mu}$ denotes the $W$ boson field, $W_{\mu\nu}$$=$$\partial_{\mu}W_{\nu}-
\partial_{\nu}W_{\mu}$, $V_{\mu\nu} = \partial_{\mu}V_{\nu}-
\partial_{\nu}V_{\mu}$, 
$V$$=$$\gamma$ or $Z$, and $M_W$ is the mass of the $W$ boson.  The global coupling parameters $g_{WWV}$ are $g_{WW\gamma}$$=$$-e$ and $g_{WWZ} = -e\, {\rm cot} \theta _{W}$, as in the standard model (SM) in which $e$ and $\theta_W$ are the magnitude of the electron charge and the weak mixing angle, respectively.
In the SM $\lambda_\gamma$$=$$\lambda_Z$$=$$0$ and $g_1^{\gamma}$$=$$g_1^Z$$=
$$\kappa_{\gamma}$$=$$\kappa_Z$$=$$1$. For convenience, anomalous trilinear gauge couplings (anomalous TGCs) $\Delta\kappa_V$ and $\Delta g_1^Z$ are defined as $\kappa_V-1$ and $g_1^Z-1$, respectively. 

The $W$ boson magnetic dipole $\mu_W$ and electric quadrupole $q_W$ moments may be expressed in terms of the coupling parameters as
\begin{equation}
\mu_W = \frac{e}{2M_W} (g^\gamma_1 + \kappa_\gamma + \lambda_\gamma) \nonumber
\end{equation}
\begin{equation}
q_W = - \frac{e}{M^2_W} (\kappa_\gamma - \lambda_\gamma) \nonumber
\end{equation}
\noindent As mentioned above, $g^\gamma_1$$=$$1$. 

If the coupling parameters have non-SM values then the amplitudes for gauge boson pair production grow with energy, eventually violating tree-level unitarity.  The unitarity violation can be controlled by parametrizing the anomalous couplings as dipole form factors with a cutoff scale, $\Lambda$.  The anomalous couplings then take a form $a(\hat{s}) = a_0 / (1 + \hat{s}/\Lambda^{2})^{2}$ in which $\sqrt{\hat{s}}$ is the center-of-mass energy of the colliding partons and $a_0$ is the coupling value in the  limit $\hat{s} \rightarrow 0$~\cite{newref}.  The quantity $\Lambda$ is physically interpreted as the mass scale where the new phenomenon responsible for the anomalous couplings is directly observable. The cutoff $\Lambda$ is conservatively set at the limit of sensitivity, close to the collision center-of-mass energy. We use $\Lambda=2$~TeV; coupling limits depend only weakly on $\Lambda$ for $\Lambda>1$~TeV in hadronic collisions at Tevatron energies.
              
\begin{figure}
\includegraphics[width=3in]{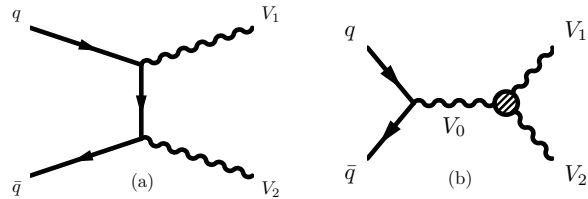}
\caption{\label{fig-feyn}
 Vector boson pair production via (a) $t$-channel and (b) $s$-channel diagrams.
For $V_{1}=W$ and $V_{2}=\gamma/Z$, $V_{0} = W$.
 For $V_{1}=V_{2}=W$, $V_{0} = \gamma/Z$.
}
\end{figure}

We measure the electroweak coupling parameters through the study of gauge boson pairs. Several processes contribute to SM boson pair production.  Fig.~\ref{fig-feyn}(a) shows $t$-channel production of dibosons in which $V_1V_2$ are $WW$, $WZ$, or $W\gamma$. The $s$-channel production shown in Fig.~\ref{fig-feyn}(b) involves boson self-interactions through a trilinear gauge vertex. Final states ($V_1V_2$) produced via the $WWZ$ coupling are $WW$ or $WZ$. Final states produced through the $WW\gamma$ coupling are $WW$ or $W\gamma$. The typical effect of anomalous TGCs is to increase the cross section especially at high boson transverse momentum ($p_T$). We thus analyze corresponding observables to measure such effects. 

Previously published limits on anomalous TGCs from a combination of channels come from the D0 Collaboration in the 1992-1996 Tevatron run with integrated luminosity ($\cal{L}$) of 100~pb$^{-1}$~\cite{run1prd}, the CDF Collaboration with the current Tevatron run ($\cal{L}$$\sim$$350$~pb$^{-1}$)~\cite{cdfresult}, and LEP2 experiments~\cite{LEP}. The best previously published $W$ boson magnetic dipole moment result is from a combination of measurements by the DELPHI Collaboration~\cite{delphi}.

In this Letter, we investigate the $WW\gamma$ and $WWZ$ trilinear vertices through diboson production. We set limits on the non-SM or anomalous TGC parameters $\lambda_V$, $\Delta\kappa_V$, and $\Delta g_1^Z$. These limits are derived from a combination of previously published measurements involving four final states: $W\gamma$$\to$$\ell \nu \gamma$~\cite{wgamprl}, $WW$$\rightarrow$$\ell \nu \ell^\prime \nu$~\cite{wwpaper}, $WZ$$\rightarrow$$\ell \nu \ell^\prime \bar{\ell}^\prime$~\cite{wzpaper}, and $WW/WZ$$\rightarrow$$\ell\nu j j$~\cite{wwwzpaper}, in which $\ell$ is an electron or muon, $\nu$ is a neutrino, and $j$ is a jet. Each measurement used data collected by the D0 detector \cite{dzeronim} from $p\bar{p}$ collisions at $\sqrt{s}=1.96$~TeV delivered by the
Fermilab Tevatron Collider.

The process $W\gamma$$\rightarrow$$\ell \nu \gamma$ is sensitive only to the $WW\gamma$ coupling. The process was studied with data corresponding to 0.7~fb$^{-1}$~\cite{wgamprl}. The main requirements were an electron with transverse energy $E_T$$>$25~GeV or a muon with transverse momentum $p_T$$>$20~GeV, a photon with $E_T$$>$9~GeV, missing transverse energy \met$>$25 (20)~GeV for the electron (muon) channel, and separation between the photon and lepton in $\eta$$-$$\phi$ \cite{measures} space of $\Delta R$$=$$\sqrt{ (\Delta \eta)^2+(\Delta \phi)^2}$$>$$0.7$. Furthermore, to suppress final state radiation the three-body transverse mass \cite{mt3} of the lepton, photon, and \met\ was required to exceed 120 (110)~GeV for the electron (muon) channel. In total 180 (83) candidate $e\nu\gamma$ $(\mu\nu\gamma)$ events were observed. After subtracting backgrounds, the signal was $130 \pm 14_{\text{stat}}$$\pm$$3.4_{\text{syst}}$ $(57
$$\pm$$8.8$$\pm$$1.8)$ events, consistent with the SM prediction of $120$$\pm$$12$ $(77$$\pm$$9.4)$ events for the $e\nu\gamma$ $(\mu\nu\gamma)$ channel. The photon $E_T$ spectra of the $W\gamma$ candidates in the data and those estimated for the backgrounds are input into the combination. For $W\gamma$ production in the presence of TGCs, spectra were simulated using the Baur Monte Carlo (MC) \cite{baurwgamlo, baurwgamnlo} with a fine grid in $\lambda_\gamma$$-$$\Delta\kappa_\gamma$ space. 

The $WW$$\to$$\ell\nu\ell^{\prime}\nu$ measurement \cite{wwpaper} used data corresponding to an integrated luminosity of 1~fb$^{-1}$.  The data were divided into three channels defined by the flavor of the leptons from the $W$ boson decays: $ee$, $e\mu$, and $\mu\mu$. For all channels, the leading lepton had $p_T$$>$25~GeV and the trailing lepton had $p_T$$>$15~GeV. The leptons were required to have opposite charge. In the data 22 ($ee$), 64 ($e\mu$) and 14 ($\mu\mu$) candidate events were observed, consistent with the sum of SM $WW$ and backgrounds of $23.5$$\pm$$1.9$ ($ee$), $68.6
$$\pm$$ 3.9$ ($e\mu$) and $10.8$$ \pm$$ 0.6$ ($\mu\mu$) events. Two-dimensional histograms of leading and trailing lepton $p_T$ were produced for the data and backgrounds and used as inputs in the combination. Distributions for SM and anomalous TGC values were generated using the $WW/WZ$ event generator from Hagiwara, Zeppenfeld, and Woodside (HZW) \cite{HWZ}. 

The $WZ$$\rightarrow$$\ell \nu \ell^\prime \bar{\ell}^\prime$ measurement \cite{wzpaper} selected the four final states $eee$, $ee\mu$, $\mu\mu e$, and $\mu\mu\mu$. The data corresponded to an integrated luminosity of 1~fb$^{-1}$. All three charged leptons were required to have $p_T$$>$$15$~GeV. $Z$ boson candidates consisted of like-flavor lepton pairs with mass $71$$<$$M_{ee}$$<$$111$~GeV or $50$$<$$M_{\mu\mu}$$<$$130$~GeV. For the $eee$ and $\mu\mu\mu$ channels, the oppositely charged lepton pair with mass closest to the $Z$ pole mass was chosen as the $Z$ boson candidate. To select $W$ boson candidates, the \met\ must have exceeded 20~GeV. To reduce background events from $t\bar{t}$ to a negligible level, the magnitude of the vector sum of the charged lepton transverse momenta and the \met\ was required to be less than 50~GeV. The sum over all channels yielded 13 candidate events in the data consistent with a SM estimate of $9.2 \pm 1.0$ $WZ$ events and $4.5 \pm 0.6$ background events. The $p_T$ of the $Z$ boson is sensitive to anomalous TGCs and is used in the combination. The HZW MC is used to estimate the SM spectrum as well as spectra from anomalous TGCs.

Finally, the $WW/WZ$$\to$$\ell \nu j j$ measurement \cite{wwwzpaper} selected events in which one $W$ boson decays leptonically and the other boson decays hadronically. The data corresponded to an integrated luminosity of $1.1$~fb$^{-1}$. The main requirements were an electron or muon with $p_T$$>$20~GeV, \met $>$20~GeV, and at least two jets with $p_T$$>$20~GeV with the leading jet satisfying $p_T$$>$30~GeV. In total 12,473 (14,392) candidate events in the $e\nu jj$ $(\mu \nu jj)$ channel were observed, consistent with the SM prediction of 12,460$\pm$550 (14,370$\pm$620) $e\nu jj$ $(\mu \nu jj)$ events \cite{lnujjcaveat}. An observable sensitive to anomalous TGCs is the $p_T$ of the dijet system. The data and background spectra for this variable are used as inputs for the combination. Spectra with anomalous TGCs were generated with the HZW MC. 

Distributions of the sensitive observables mentioned above for each final state are generated for signal with the corresponding Monte Carlos and for backgrounds using simulations or data. The signal distributions vary as a function of the TGC parameters under study both in spectral shape and event yield.  In addition to allowing variation in the TGC parameters themselves, nuisance parameters are used to allow systematic offsets to vary within their uncertainties. A simultaneous fit to the data distributions is performed in order to determine the anomalous TGC limits. The $\chi^2$ function used in this fit is~\cite{wadeLim}:

\begin{eqnarray}
\label{eqn:dchi2}
\chi^2 &=& -2\ln \left (\prod_{i=1}^{N_b}\frac{\mathcal{L}^{P}(d_i ; m_i ( \vec{R} ) ) }{\mathcal{L}^{P}(d_i; d_i)} \prod_{k=1}^{N_s}\frac{\mathcal{L}^{G}(R_k \sigma_k; 0, \sigma_k)}{\mathcal{L}^{G}(0; 0, \sigma_k)}\right ) \nonumber \\
&=&
2\sum_{i=1}^{N_b} m_i(\vec{R}) - d_i - d_i \ln\left(\frac{m_i (\vec{R})}{ d_i}\right) + \sum_{k=1}^{N_s} R_k^2,~
\end{eqnarray}

\noindent
in which the variables $i$ and $k$ index the number of histogram bins
($N_b$) and the number of systematic uncertainties ($N_s$) respectively.  In
this function ${\mathcal L}^P(\alpha;\beta)$ is the Poisson
probability for $\alpha$ events with a mean of $\beta$ events;
${\mathcal L}^G(x;\mu,\sigma)$ is the Gaussian probability for the
value $x$ in a distribution with a mean value of $\mu$ and a variance
$\sigma^2$; $R_k$ (in vector form as $\vec{R}$) is a dimensionless parameter describing departures
in nuisance parameters in units of the associated systematic
uncertainty $\sigma_k$; $d_i$ is the number of data events in bin $i$;
and $m_i(\vec{R})$ is the number of predicted events in bin $i$.  The
number of bins used in the fit is the sum of the number of bins in
each kinematic distribution for each channel.

In total 49 sources of systematic uncertainty are considered.  As implied in Eq.~\ref{eqn:dchi2}, systematic uncertainties are treated as Gaussian priors on the expected number(s) of events.  Systematic uncertainties on the luminosity, lepton identification, and theoretical uncertainties on the cross sections for the backgrounds estimated from MC are correlated across all observables.  Uncertainties on background estimates based on data are correlated across specific final states within a diboson production channel as appropriate.  The uncertainties with the largest impact on the result are those related to background cross sections and the luminosity. The effect of incorporating systematic uncertainties into the fit is to degrade the resulting limits by $\sim$30\%.

\begin{figure}[th] 
\includegraphics[width=3.15in]{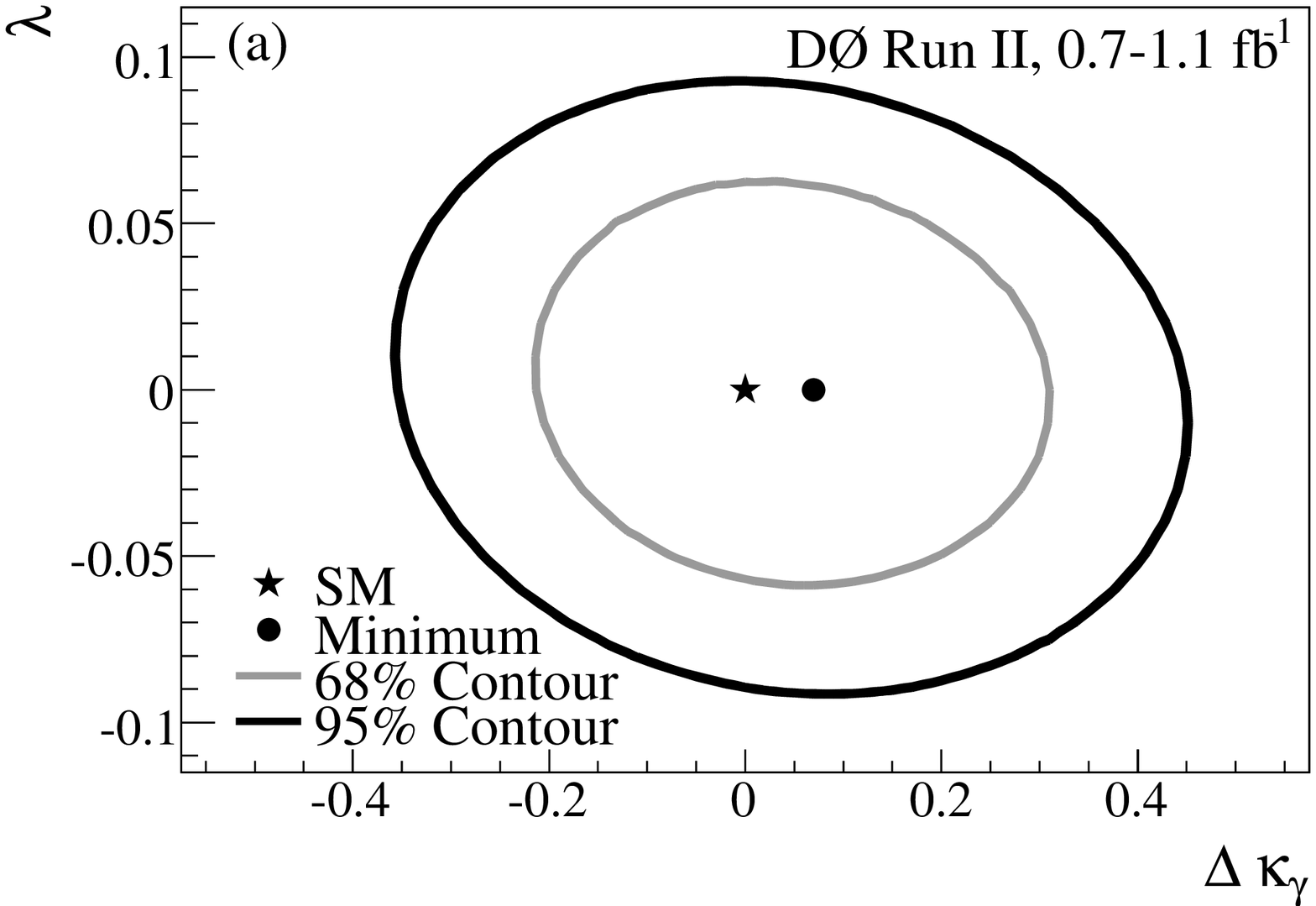} \\
\includegraphics[width=3.15in]{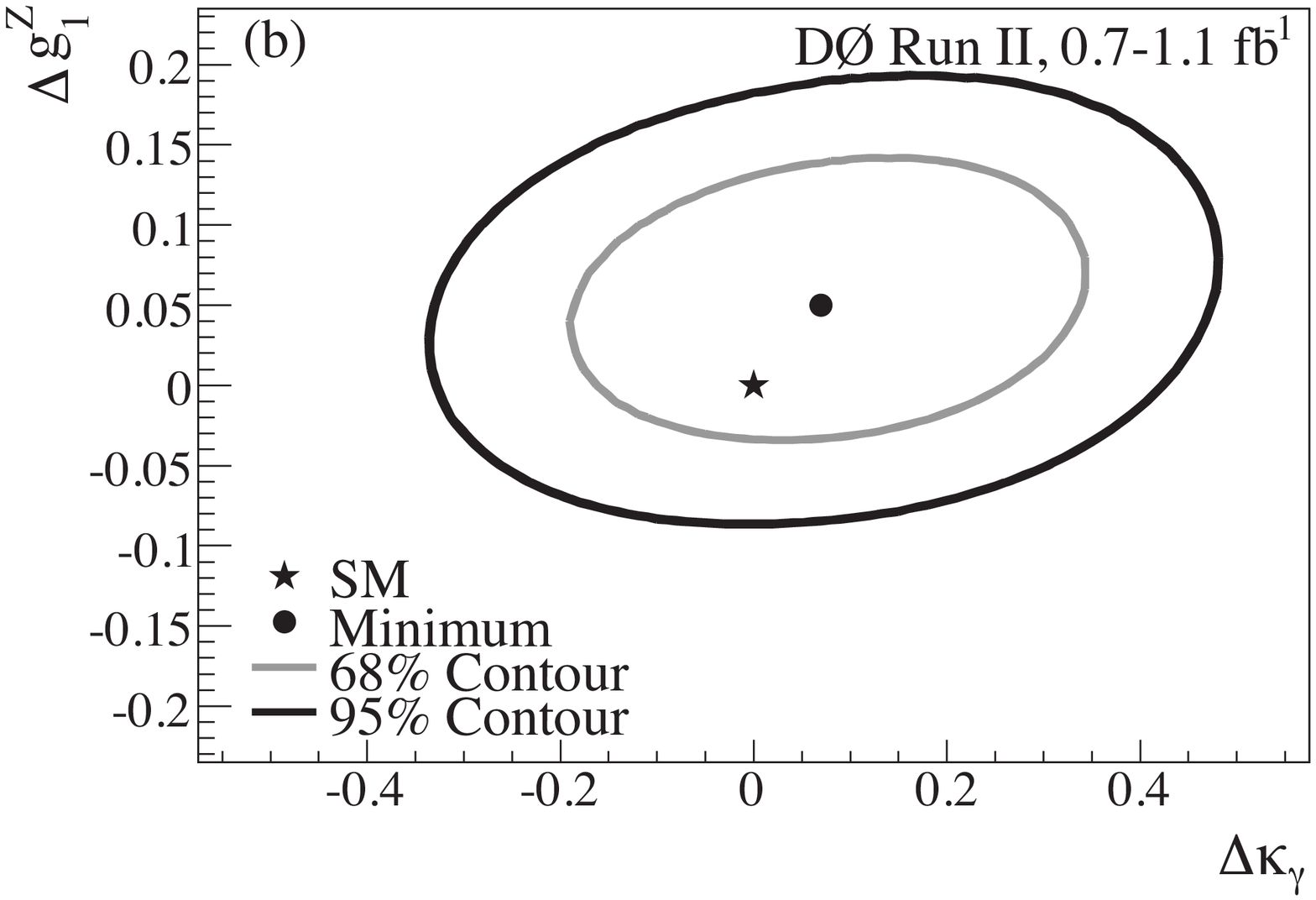} \\
\includegraphics[width=3.15in]{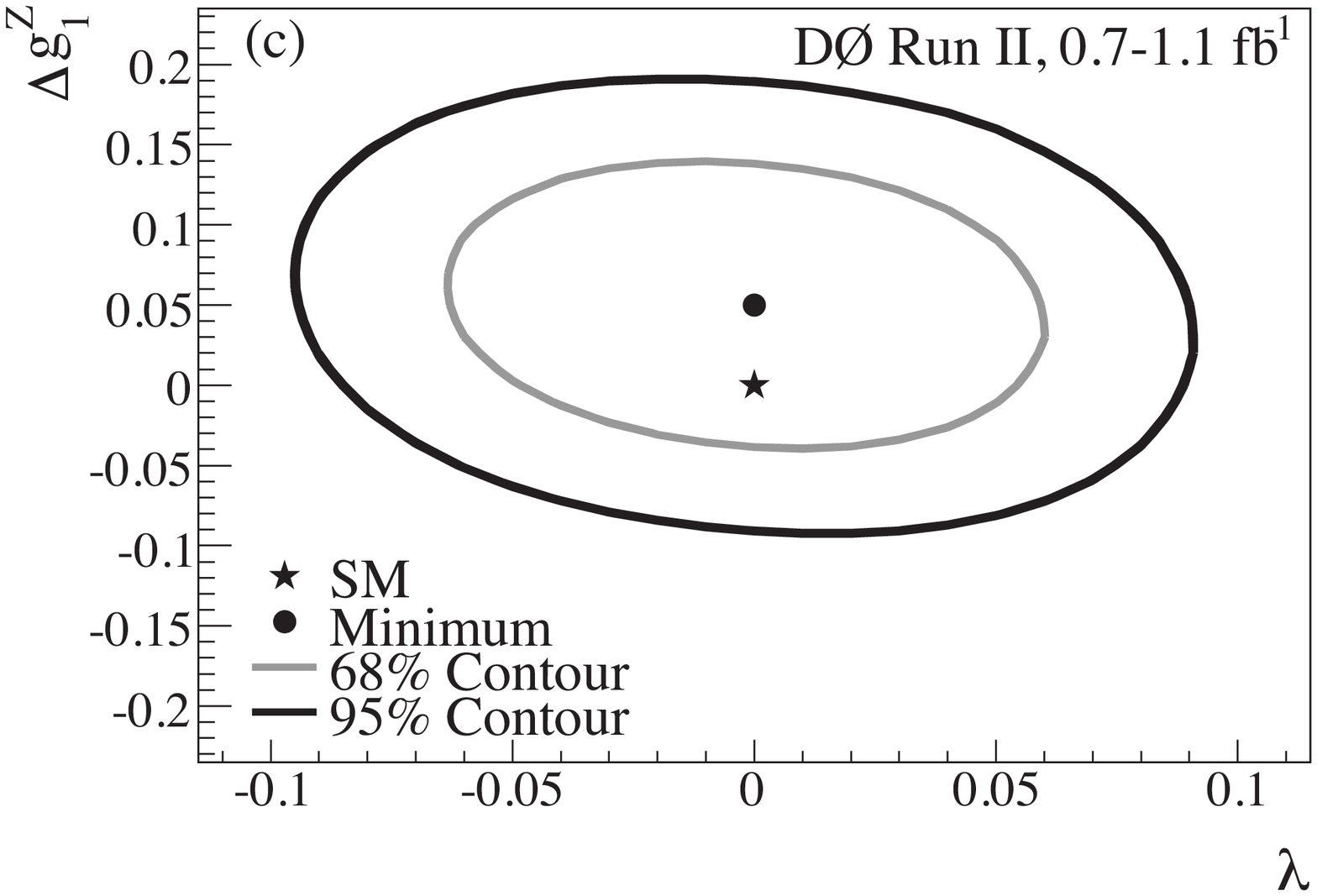}
\caption{Two-dimensional 68\% and 95\%~C.L.\ limits when respecting $SU(2)_L \otimes U(1)_Y$ symmetry and assuming $\Lambda = 2$ TeV, for (a) $\lambda$ vs.\ $\Delta\kappa_\gamma$, (b) $\Delta g_1^Z$ vs.\ $\Delta\kappa_\gamma$, and (c) $\Delta g_1^Z$ vs.\ $\lambda$.  In each case, the third 
    coupling is set to its SM value.}  \label{fig:2dlimits_1}
\end{figure}

\begin{figure}[th] 
 \includegraphics[width=3.15in]{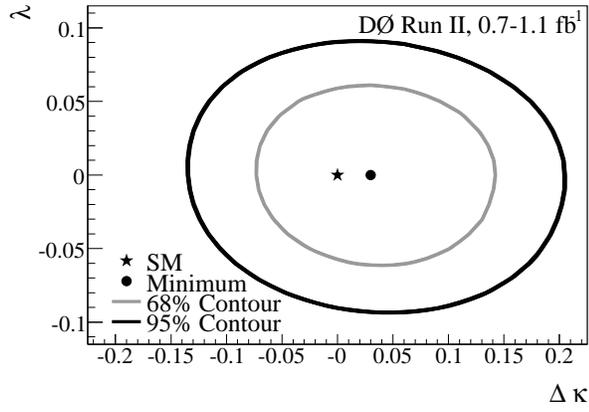}	   
   \caption{Two-dimensional 68\% and 95\%~C.L.\ limits for 
    $\lambda$ vs. $\Delta\kappa$
    when enforcing the equal-couplings constraints and assuming $\Lambda = 2$~TeV.}  \label{fig:2dlimits_2}
\end{figure}

\begin{figure}[th]
\includegraphics[width=3.15in]{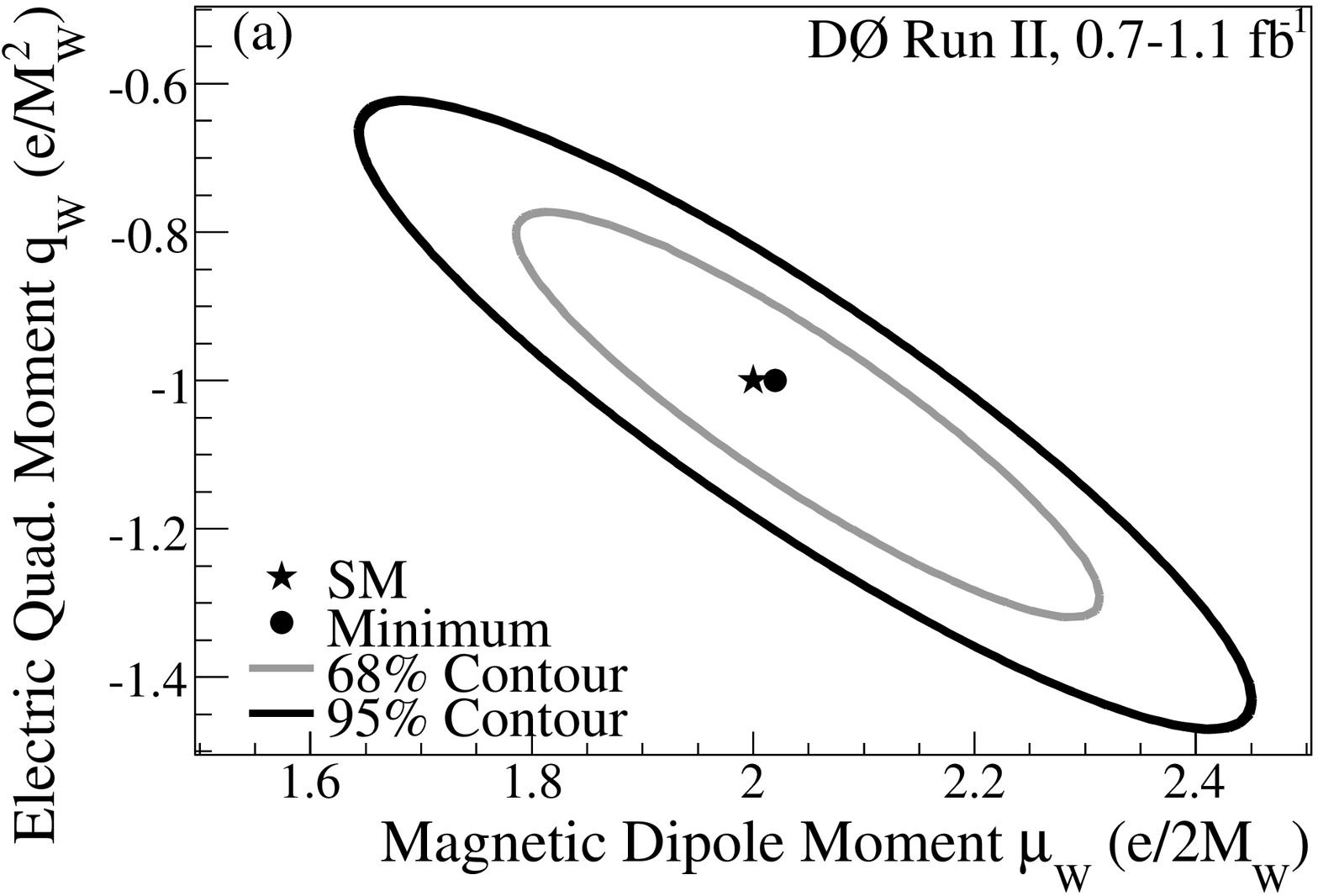}
\includegraphics[width=3.15in]{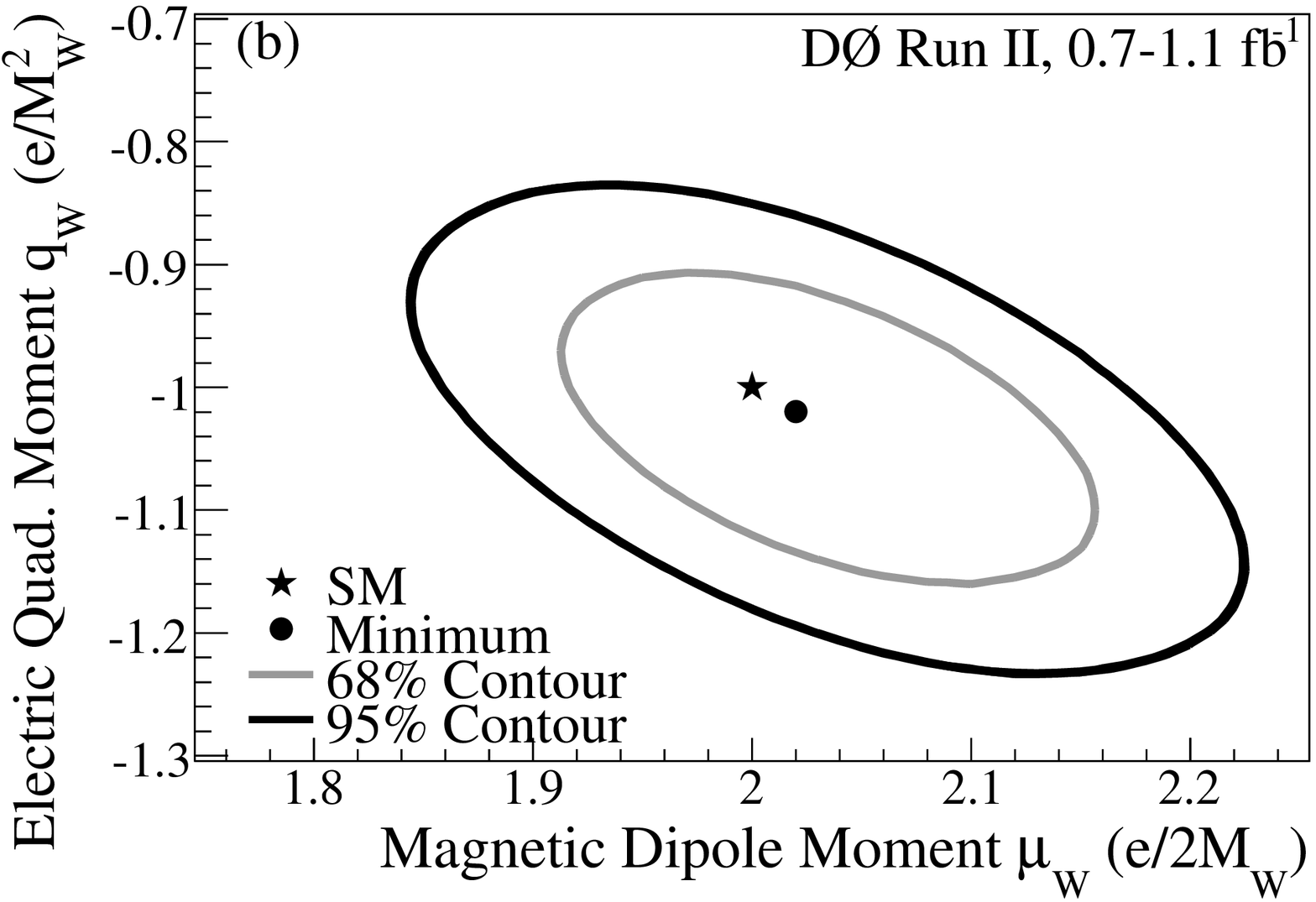}  
  \caption{Two-dimensional 68\% and 95\%~C.L. limits for the $W$ boson electric quadrupole moment vs.\ the magnetic dipole moment (a) when respecting $SU(2)_L \otimes U(1)_Y$ symmetry and (b) when enforcing equal-couplings constraints. In both cases we assume $\Lambda=2$~TeV.}
  \label{fig:2dmom}
\end{figure}

Four two-dimensional surfaces in TGC space are examined: (a) each of the
three pairings of the three free parameters
($\Delta\kappa_\gamma$, $\lambda$, $\Delta g_1^Z$) while respecting
$SU(2)_L\otimes U(1)_Y$ symmetry by using the constraints $\Delta\kappa_{Z} =
\Delta g^{Z}_{1}-\Delta\kappa_{\gamma}
\tan^{2}\theta_{W}~\text{and}~\lambda_{Z} = \lambda_{\gamma} =
\lambda$ \cite{lepscen} and (b) the $(\Delta\kappa,\lambda)$ plane for the equal-couplings scenario \cite{HWZ} in which 
$\kappa_\gamma=\kappa_Z=\kappa$, $\lambda_\gamma=\lambda_Z=\lambda$.  The two-dimensional
68\% and 95\% C.L. contours are shown in Figs.~\ref{fig:2dlimits_1}
and~\ref{fig:2dlimits_2}. The two-dimensional contours for $W$ boson magnetic dipole and electric quadrupole moments are shown in Fig.~\ref{fig:2dmom}.  The one-dimensional 68\% and 95\%
C.L. limits for each coupling parameter, with the other couplings parameters fixed at their SM values, are shown in Table~\ref{tab:1dlimits}.

\begin{table} \caption{\label{tab:1dlimits} One-dimensional $\chi^2$ minimum and 68\% and 95\% C.L. allowed intervals on anomalous values of $WW\gamma$ and $WWZ$ TGCs.  Note that $\mu_W$ and $q_W$ are in units of $(e/2M_W)$ and $(e/M^2_W)$ respectively.}
\begin{ruledtabular}
\begin{tabular}{lccc}
\multicolumn{4}{c}{Results respecting $SU(2)_L\otimes U(1)_Y$ symmetry} \\
Parameter & Minimum & 68\% C.L. & 95\% C.L.\\
\hline
$\Delta\kappa_\gamma$	& $0.07$ & $[-0.13,0.23]$ & $[-0.29,0.38]$ \\
$\Delta g_1^Z$		& $0.05$ & $[-0.01,0.11]$ & $[-0.07,0.16]$ \\
$\lambda$	                & $0.00$ & $[-0.04,0.05]$ & $[-0.08,0.08]$ \\ \\
$\mu_W$                         & $2.02$ & $[1.93,2.10]$   & $[1.86,2.16]$ \\
$q_W$                              & $-1.00$ & $[-1.09, -0.91]$ & $[-1.16, -0.84]$ \\
\hline
\multicolumn{4}{c}{Results for equal-couplings} \\
Parameter & Minimum & 68\% C.L. & 95\% C.L.\\
\hline
$\Delta\kappa$	& $0.03$&$[ -0.04,0.11]$ & $[-0.11,0.18]$ \\
$\lambda$	& $0.00$&$[-0.05,0.05]$ & $[-0.08,0.08]$ \\ \\
$\mu_W$                         & $2.02$ & $[1.94,2.09]$   & $[1.88,2.15]$ \\
$q_W$                              & $-1.02$ & $[-1.09, -0.94]$ & $[-1.16, -0.87]$ \\
\end{tabular}
\end{ruledtabular}
\end{table}

These results provide the most stringent limits on anomalous values of $WW\gamma$ and $WWZ$ TGCs measured from hadronic collisions to date. The 95\% C.L. limits in both scenarios represent an improvement relative to the previous D0 \cite{run1prd} and CDF \cite{cdfresult} results of about a factor of 3. When respecting $SU(2)_L\otimes U(1)_Y$ symmetry, our measurements with 68\% C.L. allowed intervals of $\kappa_{\gamma}=1.07^{+0.16}_{-0.20}$, $\lambda =0.00^{+0.05}_{-0.04}$ and $g_{1}^{Z}=1.05^{+0.06}_{-0.06}$ are only factors of approximately 2 -- 3 times less sensitive than the combined results from the four LEP2 experiments: $\kappa_{\gamma}=0.973^{+0.044}_{-0.045}$, $\lambda =-0.028^{+0.020}_{-0.021}$ and $g_{1}^{Z}=0.984^{+0.022}_{-0.019}$, also at 68\% C.L.~\cite{LEP}.  Furthermore, with only~1 fb$^{-1}$ of data our sensitivity is comparable to that of an individual LEP2 experiment~\cite{aleph,opal,l3,delphi}.

We also extract measurements of the $W$ boson magnetic dipole and electric quadrupole moments.  When respecting $SU(2)_L\otimes U(1)_Y$ symmetry with $g^Z_1$$=$$1$ we measure 68\% C.L. intervals (one-dimensional with the other parameter held at its SM value) of $\mu_W=2.02^{+0.08}_{-0.09} \, (e/2M_W)$ and $q_W=-1.00\pm0.09 \, (e/M^2_W)$, respectively. The most stringent previously published result is $\mu_W=2.22^{+0.20}_{-0.19} \, (e/2M_W)$ and $q_W=-1.18^{+0.27}_{-0.26} \, (e/M^2_W)$ from the DELPHI Collaboration~\cite{delphi}. 

In summary, we presented measurements of anomalous $WW\gamma$ and $WWZ$ trilinear gauge couplings and related $W$ boson magnetic dipole and electric quadrupole moments based on the combination of four diboson production and decay channels using $0.7$$-$$1.1$~fb$^{-1}$ of data collected with the D0 detector at the Fermilab Tevatron Collider.  While many of the measurements considered in this combination are limited by statistics, projections indicate that a combination of CDF and D0 data with 5~fb$^{-1}$ each will improve the sensitivity to levels comparable or better than the combined LEP2 limits.

%
We thank the staffs at Fermilab and collaborating institutions, 
and acknowledge support from the 
DOE and NSF (USA);
CEA and CNRS/IN2P3 (France);
FASI, Rosatom and RFBR (Russia);
CNPq, FAPERJ, FAPESP and FUNDUNESP (Brazil);
DAE and DST (India);
Colciencias (Colombia);
CONACyT (Mexico);
KRF and KOSEF (Korea);
CONICET and UBACyT (Argentina);
FOM (The Netherlands);
STFC and the Royal Society (United Kingdom);
MSMT and GACR (Czech Republic);
CRC Program, CFI, NSERC and WestGrid Project (Canada);
BMBF and DFG (Germany);
SFI (Ireland);
The Swedish Research Council (Sweden);
CAS and CNSF (China);
and the
Alexander von Humboldt Foundation (Germany).
%

\begin{thebibliography}{99}
\bibitem{newref} Limits on anomalous couplings presented in this Letter
 are given as the low energy limits of the couplings.
\bibitem{measures} D0 uses a cylindrical coordinate system with the
    $z$ axis running along the beam axis. Angles $\theta$ and $\phi$
    are the polar and azimuthal angles, respectively. Pseudorapidity
    is defined as $\eta = -\ln \left[ \tan(\theta/2) \right]$ in which 
    $\theta$ is measured with respect to the primary vertex. In the
    massless limit, $\eta$ is equivalent to the rapidity $y = (1/2)
    \ln \left[ (E+p_z)/(E-p_z) \right]$. $\eta_{\rm det}$ is the
    pseudorapidity measured with respect to the center of the
    detector.
\bibitem{lnujjcaveat} Note that the backgrounds for the $WW/WZ \to \ell \nu j j$ measurement, $W$+jets, $Z$+jets, $t\bar{t}$ and single top, were all modeled with Monte Carlo simulations. The dominant $W$+jets background was scaled to match the data, while all other backgrounds were normalized using SM next-to-leading order or next-to-next-to-leading-order predictions for cross sections.
\end{thebibliography}
 \end{document}